\documentstyle[aps,epsfig]{revtex}

\begin{document}

\title{Sound emission due to superfluid vortex reconnections}

\author{M. Leadbeater$^1$, T. Winiecki$^1$, D. C. Samuels$^2$, 
C. F. Barenghi$^{2,3}$, and C. S. Adams$^1$}

\address{$^1$Department of Physics, University of Durham , Durham DH1 3LE.\\
$^2$Department of Mathematics, University of Newcastle, Newcastle NE1 7RU.\\
$^3$Isaac Newton Institute for Mathematical Sciences, University of Cambridge,
Cambridge CB3 0EH.}

\date{\today}
\draft
\twocolumn
\maketitle

\begin{abstract}

By performing numerical simulations of superfluid vortex ring collisions 
we make direct quantitative measurements of the 
sound energy released due to vortex reconnections. We show that the 
energy radiated expressed in terms of the loss of vortex
line length is a simple function of the reconnection angle.
In addition, we study the temporal and spatial distribution of the 
radiation and show that energy is emitted in the form
of a rarefaction pulse. The pulse evolves into a sound wave
with a wavelength of $6-8$ healing lengths.

\end{abstract}


The decay of superfluid turbulence in the limit of low 
temperature raises some fundamental questions in the field of quantum fluids. 
For example, vortex tangles produced in superfluid helium
at $T<100$~mK are observed to decay \cite{hend94}, but at these 
temperatures the frictional dissipation due to thermal excitations is 
practically negligible. Recent theoretical work \cite{nore97,samu98,vine00} 
has highlighted the possible role of sound emission as a dissipation mechanism.
Sound emission may occur due to vortex motion or reconnection,
however, conventional numerical simulations 
based on vortex filaments governed by incompressible
Euler dynamics (the Biot-Savart law) are unable to describe this process. 
A useful tool to study vortex-sound interactions
is the Gross-Pitaevskii (GP) equation. Although unable
to fully represent the physics of HeII, the GP equation does provide  
a sophisticated fluid dynamical model capable of 
describing vortex nucleation \cite{fris92} and reconnections 
\cite{kopl93,kopl96,tsub00}. Furthermore, the GP equation
has been shown to provide an accurate description of the
recently discovered atomic Bose-Einstein condensates \cite{fermi},
where similar issues of vortex-sound interactions are likely to occur, 
particularly in experiments where many 
vortices are formed \cite{rama99,madi00}.

In this letter, we compute the sound energy radiated 
due to superfluid vortex reconnections using the GP model.  
We parameterise the energy in terms of vortex line length, and show 
that the vortex line length destroyed during a reconnection is a simple function of the 
reconnection angle. In addition,
we measure the temporal and spatial distribution of the 
radiation and show that energy is emitted in the form
of a rarefaction pulse which subsequently disperses into
sound waves.

To produce a reconnection we collide two vortex rings whose
axes of propagation are offset. This system
has the advantage that by varying the offset
we can study a range of reconnection angles.  
In addition,
a well behaved initial state can be constructed within a spatially
confined region. We adopt the same numerical methods as in our previous work \cite{wini99,wini99b,wini00}. The key points are that
throughout the paper we use dimensionless units, where distance and velocity are 
measured in terms of the healing
length, $\xi$, and the sound speed, $c$, respectively. In addition, the asymptotic number density, $n_0$, is rescaled to unity. The initial state is taken as the product of 
two vortex ring states, 
$\psi(\mbox{\boldmath$r$},0)=\phi_{1}(\mbox{\boldmath$r-r$}_1)
\phi^*_{2}(\mbox{\boldmath$r-r$}_2)$, where $\phi_{1,2}(\mbox{\boldmath$r$})$ are
time-independent vortex ring solutions of the uniform flow equation 
found by Newton's method \cite{wini99b}. We arrange for the 
rings to propagate in the $\pm x$ directions with an offset in the
$y$ direction, i.e., 
$\mbox{\boldmath$r$}_{1,2}=(\pm 25v,\pm D/2,0)$, where $x$, $y$, and $z$
correspond to the axes shown in Fig.~1, 
$v$ is the ring velocity, and $D$ is the offset between the propagation
axes of the two rings. The initial 
state is evolved according to the dimensionless GP equation,
\begin{equation}
i\partial_t\psi=-\textstyle{1\over 2}\nabla^2\psi+(\vert\psi\vert^2-1)\psi~,
\end{equation}
using a semi-implicit Crank-Nicholson time step. To model a large box, 
we map an infinite length onto the space 
$-1\leq x' \leq 1$ using $x'=x/(\vert x\vert +\zeta)$ with $\zeta=12$. 
We use a grid spacing, $\Delta x'=0.007$, and a time-step, $\Delta t=0.01$.
Simulations have been performed with vortex ring radii of $R=5.04$, 6.00 and 7.30,
which correspond to velocties $v=0.34$, 0.3, and 0.26, respectively.

A typical sequence illustrating the vortex ring collision is shown in Fig. 1. 
As the rings approach they stretch in the $yz$ plane. 
The reconnections occur along the $z$ axis at around $t=30$.
The reconnections produce two highly
elongated rings and two sound pulses which propagate outwards along the 
$\pm z$ axis. The sound pulse appears in Fig.~1 as an oval in the centre of 
the top view at $t=40$.
The stretched rings ($t=50$) rapidly shrink into two smaller vibrating
rings which move outwards. For the offset $D=4$ shown in Fig.~1, 
the outgoing rings propagate at an angle 
$\phi =\pm 56^\circ$ to the $x$ axis. The smaller radius of the outgoing rings reflects 
the energy loss due to the reconnections.
For larger offsets, the collision is less violent than the example shown,
and the scattering angle of the outgoing rings, $\phi$, is smaller.
 
To quantify the sound energy we calculate the 
energy within a measurement sphere of radius 20. 
The energy is defined relative to a uniform laminar state as
in our previous work \cite{wini99b}. The energy per ring as a function of time
for collisions between rings with radii $R=6.0$, offset by $D=2$ to
$D=8.5$, is shown in Fig. 2(inset).
If the reconnection occurs a few healing lengths along the $z$ axis
at $t\sim 30$, and the sound pulse travels outwards
at speed $c=1$, then one would expect to observe the sound leaving the 
measurement volume between $t=45$ and $t=50$. This appears as an 
energy drop in the numerical data, Fig.~2(inset).  
After the sound pulse has left the sphere one expects 
the energy to remain approximately constant until the 
outgoing vortex rings leave between $t=60$ and $t=75$. 
This energy `plateau' is well defined in  
Fig.~2(inset) for offsets between $D=4$ and
$D=8.5$, but for smaller offsets the 
outgoing rings are too fast to be resolved from the 
sound pulse. We define the radiated sound energy as the
difference between the initial energy and the value at the centre of the plateau.
 
\begin{figure}[hbt]
\centering
\epsfig{file=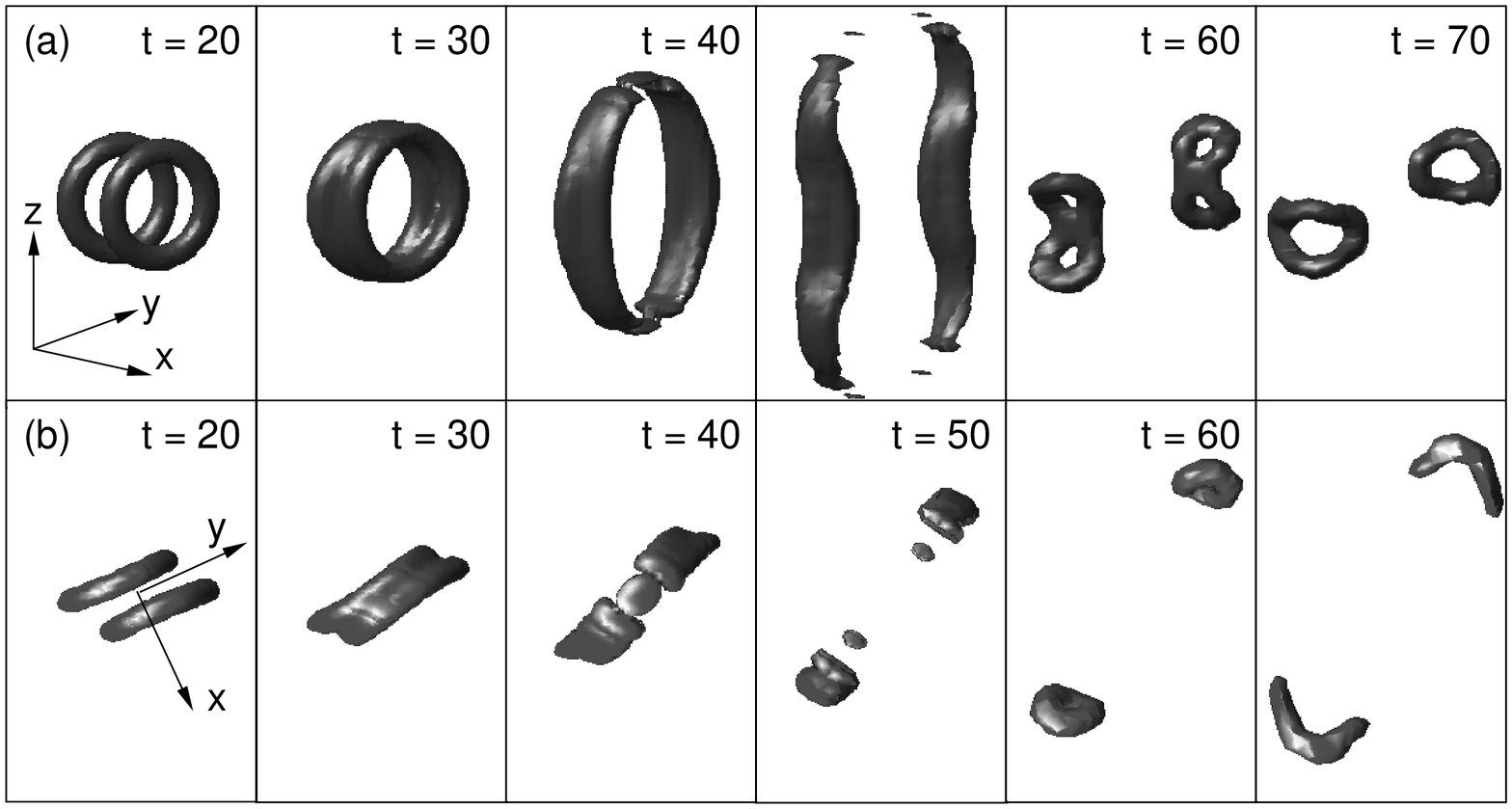,width=4.8cm,angle=-90,clip=,bbllx=35,bblly=80,bburx=430,bbury=770}
\caption{Sequence of density isosurfaces ($\vert\psi\vert^2=0.75$) 
illustrating a vortex ring collision with ring radius $R=6$ and offset $D=4$.
A side view (a) and a top view (b) are shown. Initially, the rings propagate along 
the $x$ axis and collide in the $yz$ plane at
$t\sim 30$. The collision produces highly stretched rings ($t=50$) which snap back
into two smaller rings moving at $\pm 56^\circ$ to the 
$x$ axis. The sound pulse is emitted along the $\pm z$ axis, 
and appears as an oval in the centre of the top view at $t=40$.}
\label{fig:1}
\end{figure}

A convenient way to express the energy loss is in terms of an effective vortex line 
length destroyed. To calculate the change in vortex line length we convert between
energy and length using the analytical result for the
energy of a vortex ring with radius $R$ \cite{rob71},
\begin{equation} 
E=2\pi^2 R\left[\ln\left(\frac{8R}{a}\right)-1.615\right]~,
\label{eq:e_ring}
\end{equation}
with $a=1/\sqrt{2}$. Clearly, this is approximate as the outgoing vortex
rings are excited. The vortex line length destroyed, $l_{\rm loss}$, 
as a function of reconnection angle, 
$\theta=2\cos^{-1}(D/2R)$, is plotted in Fig.~2. 
For intermediate angles, the data appear to lie on a unique curve, independent of 
the initial ring radius. A good fit is given by, 
$l_{\rm loss}=\tan^2(\theta/2)=(4R^2-D^2)/D^2$.
For large $\theta$ (small offset $D$) most of the energy is emitted as sound and the 
energy plateau is not well defined leading to a large error. Additional errors occur 
when comparing Eq.~(\ref{eq:e_ring})
with our numerical energy data due to the finite size of our measurement volume.
This error is largest when the incoming and outgoing rings are largest,
i.e. for large $R$ and large $D$. This leads
to the over estimate of vortex line loss apparent for the $R=7.30$, $\theta\leq\pi/2$
data in Fig.~2. 

\begin{figure}[hbt]
\centering
\epsfig{file=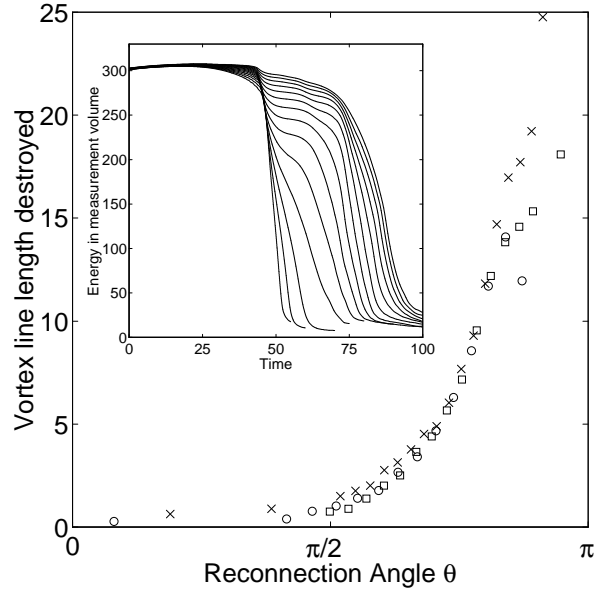,width=8cm,clip=,bbllx=65,bblly=190,bburx=570,bbury=700}
\caption{Vortex line length destroyed as a function of the reconnection angle. 
The data correspond to ring radii of $R=5.04$ 
($\circ$), 6.00 ($\Box$), and 7.30 ($\times$). The vortex line length destroyed
is calculated from the energy decrease within a measurement sphere 
when the sound pulse exits. The time-dependence of the energy 
per ring within the measurement sphere for ring  radii $R=6.0$ 
and offsets $D=2-8.5$ in increments of 0.5 is shown inset.
The sound energy exits between $t=48$ and $t=52$. The energy plateau
at $t\sim 60$ indicates the energy loss due to sound emission.}
\label{fig:2}
\end{figure}
 
The dependence of the vortex line loss on reconnection angle can in principle
be applied to determine the sound energy emitted in other reconnection geometries.
However, in the case of straight line vortices, it is known that
the lines tend to become more anti-parallel as they approach \cite{schw85,dewa94}. 
This will tend to increase the energy radiated. To convert the dimensionless 
units into values applicable to HeII, we take the number density as $n_0=2.18\times10^{28}$~m$^{-3}$, the quantum of circulation as 
$\kappa=h/m=9.98\times10^{-8}~{\rm m}^2{\rm s}^{-1}$,
and the healing length as $\xi/\sqrt{2}=0.128$~nm \cite{ray64},
then the unit of energy is $\hbar n_0c\xi^2=6.56\times10^{-24}$~J or 0.475~K.
For these parameters, the energy radiated for reconnection angles of 
$\pi/2$ and $3\pi/4$ are  
$9.2\times10^{-23}$~J (6.65~K) and $5.2\times10^{-22}$~J (37.5~K), respectively.

To determine the character of the sound radiation we have studied the 
temporal and spatial distribution of the emission.
The density along the $z$-axis for a collision with ring radius $R=6$ and offset
$D=8$ is shown in Fig.~3. Initially
the density is uniform except for a slight increase near the origin
indicating the approaching rings. Between $t=25$ and $t=30$ the rings
collide in the $xy$ plane. During the collision, the vortex cores 
merge while the rings grow outwards. Due to this stretching effect, the
reconnection point is pushed outwards along the 
$z$ axis to $z\sim\pm 7$ and delayed until $t\sim 29$ (if there were 
no distortion of the incoming rings, one
would expect the reconnection to occur at $z=\pm 4.5$ and $t=25$).
When the circulation is 
cancelled at $t=29$, the density is zero at $z=\pm7$. This density minimum 
continues to move outwards in the same direction
as the vortex cores prior to the reconnection. As the pulse moves
outwards the density mimimum gradually fills in. 

\begin{figure}[hbt]
\centering
\epsfig{file=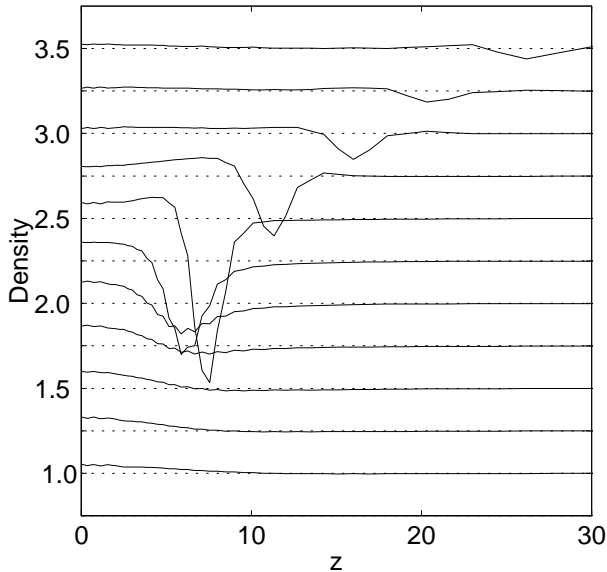,width=8cm,clip=,bbllx=105,bblly=175,bburx=520,bbury=580}
\caption{The density along the $z$ axis for a collision between two vortex
rings with radius $R=6$ and offset $D=8$. The eleven curves, corresponding to times
$t=0$ to $t=50$ in increments of 5, are offset in increments of
0.25 along the $y$ axis. Initially
the density is uniform except for a slight increase near the origin
indicating the approaching rings. Between $t=25$ and $t=30$ the rings
collide in the $xy$ plane. This results in the formation of a rarefaction
pulse which moves outwards. Note that initially the 
density at the center of the rarefaction pulse is zero and then increases.
For $t>40$ the pulse evolves into a sound wave with a wavelength of 
$6-8$ healing lengths.}
\label{fig:3}
\end{figure}

As shown for the case of vortex nucleation by a moving sphere \cite{wini00}, 
further insight can be gained by considering the time evolution as a transition
between time-independent states. For the head-on collision ($D=0$),
one can regard the segments of the ring as anti-parallel vortex lines.
The two dimensional time-independent solutions for two opposite sign vortices 
have been studied by Jones and Roberts \cite{jon82}. They show that 
when the vortex cores merge they form a rarefaction pulse. 
In a three dimensional situation, as the rarefaction pulse 
expands outwards the energy per unit length decreases which corresponds
to a lower energy on the dispersion curve. 
The lower energy rarefaction pulse has a higher central density 
and moves faster eventually approaching the sound speed \cite{jon82}. This 
exactly describes the behaviour apparent in Fig.~3.
Eventually the rarefaction pulse evolves into a sound
wave with a central wavelength of approximately 7 healing lengths. 
For HeII, taking $\xi/\sqrt{2}=0.128$~nm \cite{ray64} this converts to 1.3~nm, which corresponds to an intermediate phonon wavelength  (cf. maxon and roton wavelengths of 
0.6~nm and 0.3~nm, respectively)

The spatial extent of the sound pulse in the $x=0$ plane for the same parameters 
as in Fig.~3 but at a later time, $t=54$, is shown in Fig.~4. The density 
variation around the origin is a remnant of the departing vortex rings. The 
sound pulse appears as two crescent shaped density waves with an
angular distribution roughly equal to the reconnection angle 
(indicated by the white lines in Fig.~4).
A similar distribution is observed in the $y=0$ plane.
By $t=54$, the sound energy is spread over a large
area and the amplitude of the density variation is very small
(only a few percent). The spherical 
wave fronts are consistent with the reconnection position
and time indicated by Fig.~3.

\begin{figure}[hbt]
\centering
\epsfig{file=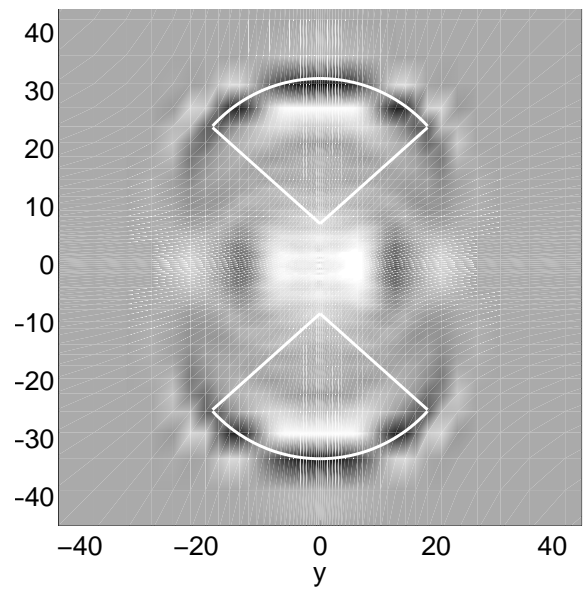,width=8cm,angle=0,clip=,bbllx=110,bblly=170,bburx=515,bbury=575}
\caption{Density cross-sections in the $x=0$ plane at $t=54$ for ring radii $R=6$ with offset $D=8$ (same parameters as in Fig.~3). The sound pulse appears
as two arcs with radius of curvature, 25, suggesting that the reconnections occurred at
$z=\pm7$ and $t=29$ consistent with Fig.~3. The angular
spread is approximately equal to the reconnection angle ($\theta=96^\circ$
for this example). The white lines indicate the positions of the reconnections, 
$z=\pm7$, the reconnection angle, $\theta$, and the expected position of the sound pulse. 
Grey scale: black 0.95; white 1.025.}
\label{fig:4}
\end{figure}

In summary, we have made direct quantitative measurements of the sound energy emitted
due to superfluid vortex reconnections. We show that the 
energy radiated can be parameterised in terms of the loss of vortex
line length which is a simple function of the reconnection angle.
The energy radiated increases dramatically as the lines become
more anti-parallel which suggests that reconnections may be
a significant decay mechanism for superfluid turbulence in the limit
of low temperature. We also show that the radiation initially appears in the form
of a rarefaction pulse which evolves into a sound wave with a
wavelength of $6-8$ healing lengths.

\acknowledgements
We thank would to thank the Newton Institute, Cambridge University, where
part of this work was completed. Financial support was provided by the EPSRC.


\begin{references}

\bibitem{hend94} P. C. Hendry, N. S. Lawson, R. A. M. Lee, and
P. V. E. McClintock, Nature {\bf 368}, 315 (1994).

\bibitem{nore97} C. Nore, M. Abid, and M. E. Brachet, 
Phys. Rev. Lett. {\bf 78}, 3896 (1997).

\bibitem{samu98} D. C. Samuels and C. F. Barenghi, 
Phys. Rev. Lett. {\bf 81}, 1644 (1998).

\bibitem{vine00} W. F. Vinen, Phys. Rev. B {\bf 61}, 1410 (2000).

\bibitem{fris92} T. Frisch, Y. Pomeau, and S. Rica, 
Phys. Rev. Lett. {\bf 69}, 1644 (1992).

\bibitem{kopl93} J. Koplik and H. Levine, Phys. Rev. Lett. {\bf 71}, 1375 (1993).

\bibitem{kopl96} J. Koplik and H. Levine, Phys. Rev. Lett. {\bf 76}, 4745 (1996).

\bibitem{tsub00} M. Tsubota, S. Ogawa, and Y. Hattori, 
J. Low Temp. Phys. (to appear).

\bibitem{fermi} See e.g. {\it Bose-Einstein condensation in atomic gases}, Proc. 
Int. School of Physics Enrico Fermi, eds. M. Inguscio, S. Stringari and C.
Wieman (IOS Press, Amsterdam, 1999).

\bibitem{rama99} C. Raman, M. K\"ohl, R. Onofrio, D. S.
Durfee, C. E. Kuklewicz,  Z. Hadzibabic, and W. Ketterle, Phys. Rev. Lett.
{\bf 83}, 2502 (1999).

\bibitem{madi00} K. W. Madison, F. Chevy, W. Wohlleben, and J. Dalibard,
cond-mat/0004037.

\bibitem{wini99} T. Winiecki, J. F. McCann, and C. S. Adams,
Phys. Rev. Lett. {\bf 82}, 5186 (1999).

\bibitem{wini99b} T. Winiecki, J. F. McCann, and C. S. Adams,
Europhys. Lett. {\bf 48}, 475 (1999).

\bibitem{wini00} T. Winiecki and C. S. Adams,
cond-mat/0006125, Europhys. Lett. (to appear).

\bibitem{rob71} P. H. Roberts  and J. Grant, J. Phys. A {\bf 4}, 55 (1971).

\bibitem{jon82} C. A. Jones and P. H. Roberts, J. Phys. A {\bf 15}, 2599 (1982).

\bibitem{schw85}  K. W. Schwarz, Phys. Rev. B {\bf 31}, 5782 (1985).

\bibitem{dewa94} A. T. A. M. DeWaele and R. G. K. M. Aarts, 
Phys. Rev. Lett. {\bf 72}, 482 (1994).

\bibitem{ray64} G. W. Rayfield and F. Reif, 
Phys. Rev. {\bf 136}, 1194 (1964).


\end{references}
\end{document}